
\headline={\ifnum\pageno=1\firstheadline\else
\ifodd\pageno\rightheadline \else\leftheadline\fi\fi}
\def\firstheadline{\hfil}
\def\rightheadline{\hfil}
\def\leftheadline{\hfil}

\footline={\ifnum\pageno>0 \hss \folio \hss \else\fi}


\font\tenbf=cmbx10
\font\tenrm=cmr10

\font\elevenbf=cmbx10 scaled\magstep 1
\font\elevenrm=cmr10 scaled\magstep 1
\font\elevenit=cmti10 scaled\magstep 1

\hsize=6.0truein
\vsize=8.5truein
\parindent=1.5pc
\baselineskip=10pt
\rightline{CERN-TH.7466/94}
\rightline{McGill/94--52}
\centerline{\tenbf Spatial Geometry and the Wu-Yang
 Ambiguity}\footnote{}
{Based on talks given by R. Khuri at PASCOS-94 meeting,
Syracuse, May 1994 and
at Gursey Memorial Conference I, Istanbul, June 1994 and
condensed from
work already published [2,3].}
\vglue 1.0cm
\centerline{\tenrm Daniel Z. Freedman$^{a,b}$ and
Ramzi R. Khuri$^{a,c}$
\footnote{$^\dagger$}
{Supported by a World Laboratory Fellowship}}
\baselineskip=13pt
\centerline{\elevenit $^a$ CERN, Theory Division}
\baselineskip=12pt
\centerline{\elevenit CH-1211 Geneva 23 Switzerland}
\baselineskip=13pt
\centerline{\elevenit $^b$
Department of Applied Mathematics, MIT}
\baselineskip=12pt
\centerline{\elevenit Cambridge, MA 02139 USA}
\baselineskip=13pt
\centerline{\elevenit $^c$ McGill University, Physics Department}
\baselineskip=12pt
\centerline{\elevenit Montreal, PQ, H3A 2T8 Canada}
\vglue 0.8cm
\centerline{\tenrm ABSTRACT}
\vglue 0.3cm
{\rightskip=3pc
 \leftskip=3pc
 \tenrm\baselineskip=12pt\noindent
We display continuous families of SU(2) vector
potentials
$A_i^a(x)$ in 3 space dimensions which generate the same
magnetic
field $B^{ai}(x)$ (with det $B\neq 0$).  These Wu-Yang families
are
obtained from
the Einstein equation $R_{ij}=-2G_{ij}$ derived recently via a
local
map of the gauge field system into a spatial geometry with
$2$-tensor
$G_{ij}=B^a{}_i B^a{}_j\det B$ and connection $\Gamma_{jk}^i$
with
torsion defined from gauge covariant derivatives of $B$.
\vglue 0.6cm}
\elevenrm
\baselineskip=14pt
\def\IR{{I\kern-0.25em R}}

The Wu-Yang ambiguity [1] is the phenomenon that two or more
gauge
inequivalent non-abelian potentials $A_i^a(x)$ generate the
same
field
strength $F_{ij}^a(x)$.  Although the original example is
3-dimensional,
it was
mainly the
4-dimensional case which was of past interest.  Many examples of
a
discrete
ambiguity have been exhibited, specifically two potentials $A$
and
$\bar A$
giving the same $F$ (see [3] and references therein).
  The few examples of a continuous
ambiguity were
degenerate in some way: for example, they were effectively
2-dimensional.
In this talk we summarize previously published work
[2,3] and display
examples of
continuous families of potentials which generate the same
magnetic
field
$$
B^{ai}=\epsilon^{ijk}\left[\partial_j A_k^a+{1\over 2}
\epsilon^{abc}
A_j^b
A_k^c\right]\ .
\eqno(1)$$
In 3 dimensions there is no ``algebraic obstruction'' to an
ambiguity. However, this fact is not
sufficient to demonstrate that (1), viewed as a partial
differential
equation
for $A_j^b$ given $B^{ai}$, has multiple solutions, and it is
this
which we
wish to explore here.

The 3-dimensional case is relevant for the Hamiltonian form of
gauge
field
dynamics in $3+1$ dimensions and especially for an attempt
[2]
to transform
from $A_i^a$ to $B^{ai}$ as the basic field variables. An
intermediate step is to replace the $A$, $B$ system by a set of
gauge-invariant
spatial geometric variables, namely a metric $G_{ij}$ and
connection
$\Gamma_{ij}^k$ with torsion.  It turns out that the
information we
find on
the Wu-Yang ambiguity invalidates the proposed form of
Hamiltonian dynamics [2].  But the geometry is valid, and
it is
through the
geometrical equations that our Wu-Yang information is obtained.

 We begin with our first example.  Consider the smooth,
algebraically
non-singular ({\it i.e.\/} det $B\neq0$) magnetic field
$B^{ai}=\delta^{ai}$,
in Euclidean space with Cartesian coordinates $x,y,z$. It is
easy to show explicitly that, for any real parameter $\beta$
with
$|\beta|>1$,
the 1-parameter family of potentials
$$
A_i^a=\pmatrix{\beta\pm\sqrt{\beta^2-1}
                 \cos (z/\beta) & \pm\sqrt{\beta^2-1}\sin
                 (z/\beta) & 0 \cr
               \pm\sqrt{\beta^2-1}
                 \sin (z/\beta) & \beta\mp\sqrt{\beta^2-1}\cos
		 (z/\beta) & 0 \cr
	       0 & 0 & 1/\beta \cr}
\eqno(2)$$
all reproduce the same $B^{ai}$. Gauge inequivalence is
demonstrated
by the
fact that the invariants $B^{aj}D_i B^{ak}$ depend on $\beta$
and
$z$.
The particular magnetic field $B^{ai}=\delta^{ai}$ is
invariant
under rotations and translations of the configuration space
$\IR^3$ (the spatial
rotations must be combined with a suitably chosen SO(3) gauge
transformation,
constant in this case, as is well known).  Since (1) is also
covariant,
each such isometry which does not leave $A_i^a$ invariant
produces
another
Wu-Yang related potential.  In this way one can extend the
potentials
displayed
in (2) to a 4-parameter family, in which the wave has an
arbitrary
phase,
$z\to z-z_0$ and direction $(0,0,1)\to \hat{k}$.
(1) is covariant under diffeomorphisms.
Thus examples of the Wu-Yang ambiguity automatically extend to
entire
orbits
of the diffeomorphism group, and one can find a
diffeomorphism under
which the field $B^{ai}=\delta^{ai}$, which has infinite
energy, transforms to a configuration $B'^{a\alpha}(y)$ which
falls
sufficiently fast as $y^\alpha\to\infty$ that it has finite
energy.

 Let us now review the spatial geometry which is the
main tool
used in this work.  Let $B^a{}_i(x)$ denote the matrix inverse
of the
magnetic
field of SU(2) gauge theory, and $\det B=\det B^{ai}$, which is
gauge
invariant.  Then $G_{ij}(x)=B^a{}_i(x)B^a{}_j(x)\det B$ is a
gauge
invariant
symmetric tensor (under diffeomorphisms).  The following geometry, which
obviously uses the fact that the gauge group SU(2) is also the
tangent space
group of a 3-manifold, emerged from the physical aim of studying the
action
of the electric field on gauge invariant state functionals
$\psi[G]$.

The quantity $b_i^a=|\det B|^{1\over 2}B_i^a$ is essentially a
frame
for
$G_{ij}$.  One may apply a Yang-Mills covariant derivative and
define
a
quantity $\Gamma_{ij}^k$ as follows:
$$
D_i b_j^a=\Gamma_{ij}^k b_k^a .
\eqno(3)$$
It can be shown that $\Gamma$ is a metric compatible connection
 for $G$,
and can be
written as [2]
$$
\Gamma_{ij}^k=\dot\Gamma_{ij}^k(G)-{K_{ij}}^k
\eqno(4)$$
where $\dot\Gamma$ is the Christoffel connection and $K$ is the
contortion
tensor, which is antisymmetric in the last pair of indices,
$K_{ijk}=-K_{ikj}$.
Further manipulation of (3) leads to
$$
R_{ij}(\Gamma)=-2G_{ij}
\eqno(5)$$
which defines an Einstein geometry with torsion.  One may show
using the
second Bianchi identity of curvatures with torsion, that an
integrability
condition for (5) is that the contortion tensor is traceless,
${K_{kj}}^k=0$,
and that this is also a direct requirement of the gauge field
Bianchi
identity, $D_i B^{ai}=0$, applied to the definition (3) of
$\Gamma$.

The discussion above defines the forward map from Yang-Mills
fields
$A$ and
$B$, always related by (1), to geometric variables $G$ and
$\Gamma$
defined
by explicit local formulas above. The gauge field Ricci and
Bianchi
identities
then imply that $G$ and $\Gamma$ are related by the Einstein
condition (5)
with traceless contortion.  The fundamental reason for the
Einstein
geometry
is that the magnetic field is simultaneously the curvature (1)
of
the
gauge
connection $A$ and also essentially the frame of the spatial
geometry.

One may also ask about the inverse map from tensor $G_{ij}(x)$
and
connection
$\Gamma_{ij}^k(x)$ on $\IR^3$ to gauge fields.  Suppose that a
frame
$b^a_i$, with $\det b>0$, is constructed for $G_{ij}$ by any
standard
method, then (3) can be written out as
$$
\partial_i b_j^a-\Gamma_{ij}^k b_k^a+
\epsilon^{abc}A_i^b b_j^c=0\ .
\eqno(6)$$
This is just the ``dreibein postulate'' with $A$ essentially the spin
connection, and one can solve for $A$, obtaining
$$
A_i^a=-{1\over 2}\epsilon^{abc}b^{bj}\left(\partial_i
b_j^c-\Gamma_{ij}^k
b_k^c\right)
\eqno(7)$$
while the magnetic field is defined from the inverse frame by
$$
B^{ai}(x)=|\det G_{jk}|^{1\over 2} b^{ai} .
\eqno(8)$$
Thus given a frame one obtains the magnetic field from (8),
 while
both $b$
and $\Gamma$ are required to define the potential via (7).
 Since
the frame is
unique up to a local $SO(3)$ rotation, these maps define $A$
and $B$
uniquely up to an $SU(2)$ gauge transformation.  Furthermore,
$A$ and $B$ defined in this way satisfy the gauge
theory relation (1) if $\Gamma$ and $G$ satisfy (5).

Thus the gauge theory Wu-Yang ambiguity will appear whenever the
Einstein
equation (5) viewed as a partial differential equation for $K$,
given $G$, has
multiple solutions.  To investigate this it is useful to use the
representation
$$
K^i{}_{jk}=\epsilon_{jkn}S^{ni}{1\over |\det G|^{1/2}}
\eqno(9)$$
which automatically satisfies the antisymmetry
and tracelessness requirements if $S^{ni}$ is a
symmetric tensor.  When (5) is expanded out using (4)
and (9), one finds that the Einstein equation is equivalent to
$$
{\epsilon^{jkl}\over |\det G|^{1/2}}\dot\nabla_k
S_{li}-\left(S_k^j
S_i^k-S_k^k S_i^j\right)=\dot R_i^j+2\delta_i^j .
\eqno(10)$$
In (10) $\dot\nabla_k$ indicates a spatial covariant derivative
with
Christoffel connection $\dot\Gamma$ and $\dot R_{ij}$ is the
conventional
symmetric Ricci tensor.  The $\epsilon\dot\nabla S$
term is non-symmetric, so that (10) comprises 9 equations for
the
6
components of $S_{ij}$.  However it was shown explicitly [2]
 that
there is a Bianchi identity which imposes 3 constraints on the 9
equations, so there is
no reason to think that (10) is an overdetermined system. From
(4),
(7) and (9), $A$ can be expressed in terms of $S$ as
$$A_i^a=-{1\over 2}\epsilon^{abc}b^{bj}\dot\nabla_i b^c_j -
b^{ak}S_{ki} .
\eqno(11)$$

Our approach to the Wu-Yang ambiguity is to take an input metric
$G_{ij}(x)$ and study the solutions of (10) for the torsion
$S_{ij}(x)$.
It is not clear why this should be a simpler method than to
study
directly
whether (1) has multiple solutions for $A$, given $B$.
Perhaps it
is
because
an equation for the 6 components of $S_{ij}$ is simpler to
handle
than an
equation for the 9 components of $A_i^a$, but it may just be an
historical
accident that has led us to approach the Wu-Yang ambiguity via
the
spatial
geometry.

Before beginning to study applications of (10), it is perhaps
useful
to note
that (3) indicates that $\Gamma$ is completely determined by
first
covariant
derivatives $D_i B^{aj}$ of the magnetic field.  It then
follows from
properties of the inverse map discussed above that there is no
Wu-Yang
ambiguity for the potential $A_i^a$, if we require that both
$B$ and
$DB$  are
preserved\footnote{$^*$}{Actually it is sufficient to require
that
$D_{[i}b^a_{j]}$ is preserved, because this determines the
torsion
tensor from
(3).}. In 4 dimensional $SU(2)$ gauge theory, the field
strength and its first two covariant derivatives determine the
potential
locally uniquely.

The Wu-Yang ambiguity indicates that the potential $A_i^a(x)$
contains gauge
invariant information beyond that in the magnetic field
$B^{ai}(x)$.
Therefore the change of field variable $A_i^a\to B^{ai}$ which
 was
the basis
of the version of gauge invariant Hamiltonian dynamics presented in
[2] is
invalid.  The discrete 2:1 ambiguity envisaged there could be
handled, but it
is probably impossible to deal with a continuous ambiguity
without
serious
revision of the proposal.

 It is the tensor $G_{ij}(x)=\delta_{ij}$
that corresponds to the magnetic field
$B^{ai}=\delta^{ai}$, and it can be seen without
difficulties that the torsion solutions of (10) are
related in this simple case to the potentials of (2) by
$A^a_i(z)=-S_{ai}(z)$.
 Note that at $\beta=1$, the solution (2) reduces to
$S_{ij}=\delta_{ij}$.  We found the family of solutions (2)
by
first
linearizing about $S_{ij}=\delta_{ij}$ and using Fourier
 analysis to
find
linearized modes of wave number $k^2=1$.  This led us to
investigate
the
single variable ansatz $S_{ij}(z)$ which reduces (10) to a non
linear
system
of ordinary differential equations.  Some fiddling then led to
(2),
which is
unique within this ansatz (except for translation $z\to z-z_0$).
One can show that the only spherically
symmetric solutions of (10) for input $G_{ij}=\delta_{ij}$ are
the
solutions
$S_{ij}=\pm\delta_{ij}$. There is a heuristic
argument that
the potentials displayed in (2) together with those obtained
from
them by
translation and rotation are the only potentials for the field
$B^{ai}=\delta^{ai}$ which continuously limit to potentials
$\bar A^a_i=\delta^a_i$ with $\beta=1$ in (2). The reason is
that
one
can
show using the Fourier transform that the set of linear
perturbations
about
$\bar A^a_i$ obtained in the $\beta\to 1$ limit of our Wu-Yang
 family
are complete.

 Another set of Wu-Yang examples emerges from the
3-dimensional hyperbolic metrics
$$ ds^2={1\over c^2z^2}\left( dx^2 + dy^2 + dz^2\right)
\eqno(12) $$
for which $R_{ij}=-2c^2 G_{ij}$. One may anticipate that the
 case
$c^2=1$ is
especially simple because the right side of (10) vanishes. It
turns
out that
one can also make the $\epsilon \dot\nabla S$ and $SS$ terms
 vanish
separately.
There is no integrability constraint when $\dot\nabla_j$ is
applied
to the former condition, while the second condition implies that
$S_{ij}$ is a rank 1
dyadic matrix. With this structure in view one can easily find
that
within
the two variable ansatz $S_{ij}(z,x)$, there is the family of
solutions
$$ S_{ij}(z,x)=\delta_{i1} \delta_{j1} {1\over z}h(x)
\eqno(13) $$
which involves an arbitrary function of the variable $x$. The
solution can
be rotated by an angle $\theta$ in the $x,y$ plane to obtain
$$ \eqalignno{S_{ij}&={1\over z} h(x\cos\theta+y\sin\theta)
V_i V_j
,\cr
V_i&=(\cos\theta,\sin\theta,0). &(14)\cr} $$
We have not studied the application of the full $SO(2,1) \times
SO(2,1)$
isometry group of the metric (12), but more solutions seem
likely.
In
this
frame the magnetic field is given simply by
$B^{ai}=\delta^{ai}/z^2$
while the   gauge potential corresponding to (13) is obtained
 from
(11):
$$ A_1^1=-h(x), \quad A_2^1=-A_1^2=1/z, \eqno(15)$$
with the rest of the components vanishing. In this frame,
the magnetic
field $B^{ai}$ is singular on the plane $z=0$. It is
straightforward
to transform our
configuration to a frame in which both
the magnetic field and gauge potential are manifestly regular
over
all of
$\IR^3$ (see [3]).

When $c^2\neq 1$ the full nonlinear equations are very
difficult to
handle, so
we restrict ourselves to a perturbative expansion about the
symmetric
solution
$\bar S_{ij}=\sqrt{1-c^2}G_{ij}$ of (10) by setting
 $S_{ij}=\bar
S_{ij} +
\hat\Sigma_{ij}$. The perturbation $\hat\Sigma_{ij}$ satisfies
the
linear equation
$$ (cz) \epsilon^{jkl}\dot\nabla_k \hat\Sigma_{li} +
\sqrt{1-c^2}(\hat\Sigma_{ji} +
\hat\Sigma_{kk}\delta_{ji})=0 .\eqno(16) $$
(the placement of the $j$ index reflects the removal of the
 conformal
factor.)
The 9 equations for the 6 components of $\hat\Sigma$ cannot all
 be
independent,
and the fact that we find a consistent solution below is a
 practical
test
of the exact Bianchi identity satisfied by (10). Note that
the
$ij$
contraction of (16) immediately tells us that the trace
$\hat\Sigma_{kk}=0$.

Because of the $x$-translation symmetry of the metric (12) we
look
for a solution
of the form $\hat\Sigma_{ij}(z,x)=\Sigma_{ij}(z,k) e^{ikx}$. The
 9
equations of
(16) can be manipulated to obtain a second order differential
equation for the
component $\Sigma_{23}$
$$ \left(z^2 {d^2\over dz^2} + z {d\over dz} - k^2z^2 +
{1-c^2\over
c^2}\right)
\Sigma_{23}=0 ,\eqno(17) $$
while the other components are related to $\Sigma_{23}$ by
$$ \eqalignno{ \Sigma_{33}&={-ikc\over \sqrt{1-c^2}}z
\Sigma_{23}
,\cr
\Sigma_{13}&={c\over \sqrt{1-c^2}}z {d\over dz}\Sigma_{23} ,\cr
\Sigma_{12}&={i\over k}({d\over dz} - {1\over z})\Sigma_{23} ,\cr
\Sigma_{11}&={-ic\over k\sqrt{1-c^2}}z {d^2\over dz^2}\Sigma_{23}
,\cr
\Sigma_{22}&=-(\Sigma_{11} + \Sigma_{33}) . &(18)\cr }  $$
Note that (20) is the differential equation for Bessel functions
of
imaginary
argument $ikz$ and index $p=\sqrt{(c^2-1)/c^2}$ which is also
imaginary when
$c^2<1$ and the symmetric torsion $\bar S_{ij}$ is real.

Note that the wave number $k$ of the linear perturbation is not
restricted
in contradistinction to the flat metric where, as can be seen
from
the small
amplitude limit $\beta\to 1$ in (2), the wave number $k=1$ is
required. This
means that the general real superposition
$$ \int dk \varphi(k) \Sigma_{ij}(kz)e^{ikx} + {\rm c.c.}
\eqno(19)
$$
is also a solution, so we have the freedom of an arbitrary
function
at
the linear level. We expect that (19) can be used as the
``input''
to
the
system of differential equations determining second and
higher order
perturbative solutions of (10), and that the functional
freedom of
$\varphi(k)$
remains. Thus the qualitative picture of the torsion solutions
for
the
hyperbolic metrics for all values of $c$ is that they contain an
arbitrary
function of a single variable and the additional parametric
 freedom
obtained
from isometries. The case $c=1$ is special only because exact
solutions can be
easily obtained. A similar analysis to the above follows  for
the
case of $2+1$ product metrics, in which
 functional freedom is again  expected to persist in higher
 order
perturbative solutions.

 In summary, what we have discussed in this talk are several
examples of a
continuous
Wu-Yang ambiguity for $SU(2)$  gauge fields in 3 dimensions and
 a new
technique,
namely the Einstein space condition (10) for obtaining such
 field
configurations. It is intriguing to ask about the systematics
 of the
ambiguity;
namely what properties of the $B$-field determine the degree
 of
ambiguity in the
associated potentials $A$. Our examples provide at least a
limited
view of this
systematics. Certainly an ambiguity is generated whenever
 there is a
symmetry
transformation of $B$ which acts nontrivially on $A$, but
 this is not
enough to
explain the free parameter $\beta$ in (2), nor the arbitrary
functions such as
$F(x)$ in the example (13-15) or in the linear solutions.
 Gauge field topology does not seem to be the issue here
for two
reasons. First of all the ambiguity can be exhibited in any
 compact
subset
of the configuration space $\IR^3$. Second, if we are given
in some
gauge a
Wu-Yang family with suitable behavior at spatial infinity one
can
apply a
gauge transformation to change the topological class at will.
Of
course
one does expect that, except for singularities of the
map (1), the
degree
of ambiguity in $A$ will not change as the parameters of $B$ are
smoothly varied.
Our examples appear to be consistent with this requirement,
although
the
discrete ambiguity found when $\dot R_{ij}=0$ must be
understood as a
singular limit
of the case of non-zero curvature. It is interesting that the
Riemannian
curvature $\dot R_{ij}$ of the metric $G_{ij}$ obtained from
$B$
plays a role both in the
ease of obtaining solutions for $A$ and in the qualitative
nature of
the
ambiguity.

\vglue 0.6cm
\leftline{\elevenbf References}
\vglue 0.4cm
\medskip
\itemitem{1.} T.T. Wu and C.N. Yang, Phys. Rev. D
{\elevenbf 12} 3845 (1975).
\itemitem{2.} D.Z. Freedman, P.E. Haagensen, K. Johnson and
J.I.
Latorre, hepth 9309045 CERN-TH.7010/93, CTP \#2238  (1993).
\itemitem{3.} D.Z. Freedman and R.R. Khuri, Phys. Lett. B
{\elevenbf 329} 263 (1994).

\bye